\documentclass[singlecolumn,epjc3]{svjour3}
\smartqed  
\RequirePackage{graphicx}
\journalname{Eur. Phys. J. C}

\begin{document}

\title{Anisotropic models for compact stars}

\author{S.K. Maurya\thanksref{e1,addr1}
\and Y.K. Gupta\thanksref{e2,addr2} \and Saibal
Ray\thanksref{e3,addr3} \and Baiju Dayanandan
\thanksref{e4,addr4}.}

\thankstext{e1}{e-mail: sunil@unizwa.edu.om}
\thankstext{e2}{e-mail: kumar$001947$@gmail.com}
\thankstext{e3}{e-mail: saibal@iucaa.ernet.in}
\thankstext{e4}{e-mail: baijudayanand@yahoo.co.in}

\institute{Department of Mathematical \& Physical Sciences,
College of Arts \& Science, University of Nizwa, Nizwa, Sultanate
of Oman\label{addr1} \and Department of Mathematics, Jaypee
Institute of Information Technology University, Sector-128 Noida,
Uttar Pradesh, India\label{addr2} \and Department of Physics,
Government College of Engineering \& Ceramic Technology, Kolkata
700010, West Bengal, India\label{addr3} \and Department of
Mathematical \& Physical Sciences, College of Arts \& Science,
University of Nizwa, Nizwa, Sultanate of Oman\label{addr4}}

\date{Received: date / Accepted: date}

\maketitle

\begin{abstract}
In the present paper we obtain an anisotropic analogue of
Durgapal-Fuloria \cite{Durgapal1985} perfect fluid solution. The
methodology consists of contraction of anisotropic factor $\Delta$
by the help of both metric potentials $e^{\nu}$ and $e^{\lambda}$.
Here we consider $e^{\lambda}$ same as
Durgapal-Fuloria~\cite{Durgapal1985} whereas $e^{\nu}$ is that
given by Lake~\cite{Lake2003}. The field equations are solved by
the change of dependent variable method. The solutions set
mathematically thus obtained are compared with the physical
properties of some of the compact stars, strange star as well as
white dwarf. It is observed that all the expected physical
features are available related to stellar fluid distribution which
clearly indicate validity of the model.
\end{abstract}


\section{Introduction}

Few decades ago a new analytic relativistic model was obtained by
Durgapal and Fuloria~\cite{Durgapal1985} for superdense stars in
the framework of Einstein's General Theory of Relativity. They
showed that the model in connection to neutron star stands all the
tests of physical reality with the maximum mass $4.17~M_{\odot}$
and the surface redshift $0.63$. Very recently Gupta and
Maurya~\cite{Gupta2011} presented a class of charged analogues of
superdense star model due to Durgapal and
Fuloria~\cite{Durgapal1985} under the Einstein-Maxwell spacetimes.
The members of this class have been shown to satisfy various
physical conditions and exhibit features (i) with the maximum mass
$3.2860~M_{\odot}$ and the radius $18.3990$~km for a particular
interval of the parameter $1<K \leq 1.7300$, and (ii) with the
maximum mass $1.9672~M_{\odot}$ and the radius $15.9755$~km for
another interval of the parameter $1 <K \leq 1.1021$. Later on a
family of well behaved charged analogues of Durgapal and
Fuloria~\cite{Durgapal1985} perfect fluid exact solution was also
obtained by Murad and Fatema \cite{Murad2014} where they have
studied the Crab pulsar with radius $13.21$~km.

In a similar way we have considered a generalization of Durgapal
and Fuloria~\cite{Durgapal1985} with anisotropic fluid sphere such
that $p_r \neq p_t$, where $p_r$ and $p_t$ respectively are radial
and tangential pressures of fluid distribution. The present work
is a sequel of the paper \cite{Maurya2015} where we have developed
a general algorithm in the form of metric potential $\nu$ for all
spherically symmetric charged anisotropic solutions in connection
to compact stars. However, in the present study without
considering any anisotropic function we can develop algorithm by
the help of metric potentials only and here lies the beauty of the
investigation. Another point we would like to add here that till
now, as far as our knowledge is concerned, no alternative
anisotropic analogue of Duragapal-Fuloria~\cite{Durgapal1985}
solution is available in the literature.

In connection to anisotropy we note that it was Ruderman
\cite{Ruderman1972} who argued that the nuclear matter may have
anisotropic features at least in certain very high density ranges
($> ~ 10^{15}~gm/cm^3$) and thus the nuclear interaction can be
treated under relativistic background. Later on Bowers and Liang
\cite{Bowers1974} specifically investigated the non-negligible
effects of anisotropy on maximum equilibrium mass and surface
redshift. In this regard several recently performed anisotropic
compact star models may be consulted for further reference
~\cite{Mak2002,Mak2003a,Usov2004,Varela2010,Rahaman2010a,Rahaman2011,Rahaman2012a,Kalam2012,Bhar2015}.
We also note some special works with anisotropic aspect in the
physical system like Globular Clusters, Galactic Bulges and Dark
Halos in the Refs. \cite{Nguyen2013a,Nguyen2013b}.

As a special feature of anisotropy we note that for small radial
increase the anisotropic parameter increases. However, after
reaching a maximum in the interior of the star it becomes a
decreasing function of the radial distance as shown by Mak and
Harko \cite{Mak2004a,Mak2004b}. Obviously at the centre of the
fluid sphere the anisotropy is expected to vanish.

We would like to mention that algorithm for perfect fluid and
anisotropic uncharged fluid is already available in the literature
\cite{Lake2003,Lake2004,Herrera2008}. As for example, we note that
in his work Lake \cite{Lake2003,Lake2004} has considered an
algorithm based on the choice of a single monotone function which
generates all regular static spherically symmetric perfect as well
as anisotropic fluid solutions under the Einstein spacetimes. It
is also observed that Herrera et al. \cite{Herrera2008} have
extended the algorithm to the case of locally anisotropic fluids.
Thus we opt for an algorithm to a more general case with
anisotropic fluid distribution. However, in this context it is to
note that in the Ref. \cite{Maurya2015} we developed an algorithm
in the Einstein-Maxwell spacetimes.

The outline of the present paper can be put as follows: in Sec. 2
the Einstein field equations for anisotropic stellar source are
given whereas the general solutions are shown in Sec. 3 along with
the necessary matching condition. In Sec. 4 we represent
interesting features of the physical parameters which include
density, pressure, stability, charge, anisotropy and redshift. As
a special study we provide several data sheets in connection to
compact stars. Sec. 5 is used as a platform for some discussions
and conclusions.

\section{The Einstein field equations}

In this work we intend to study a static and spherically symmetric
matter distribution whose interior metric is given in
Schwarzschild coordinates, $x^i=(r, \theta, \phi, t)$
\cite{Tolman1939,Oppenheimer1939}
\begin{equation}
ds^2 = - e^{\lambda(r)} dr^2 - r^2(d\theta^2 + sin^2\theta d\phi^2) + e^{\nu(r)} dt^2.\label{metric1}
\end{equation}

The functions $\nu$ and $\lambda$ satisfy the Einstein field
equations,
\begin{equation}
\kappa{T^i}_j =  {R^i}_j - \frac{1}{2} R {g^i}_j .\label{field}
\end{equation}
where $\kappa = 8\pi$ is the Einstein constant with $G=1=c$ in
relativistic geometrized unit, $G$ and $c$ respectively being the
Newtonian gravitational constant and velocity of photon in vacua.

The matter within the star is assumed to be locally anisotropic
fluid in nature and consequently ${T^i}_j$ is the energy-momentum
tensor of fluid distribution defined by
\begin{equation}
{T^i}_j = [(\rho + p_r)v^iv_j - p_t{\delta^i}_j + (p_r - p_t)
\theta^i \theta_j],\label{matter}
\end{equation}
where $v^i$ is the four-velocity as
$e^{\lambda(r)/2}v^i={\delta^i}_4$, $\theta^i$ is the unit space
like vector in the direction of radial vector, $\theta^i =
e^{\lambda(r)/2}{\delta^i}_1$ is the energy density, $p_r$ is the
pressure in direction of $\theta^i$ (normal pressure) and $p_t$ is
the pressure orthogonal to $\theta_i$ (transverse or tangential
pressure).

For the spherically symmetric metric (\ref{metric1}), the Einstein
field equations may be expressed as the following system of
ordinary differential equations \cite{Dionysiou1982}
\begin{equation}
-\kappa {T^1}_1 = \frac{{\nu}^{\prime}}{r} e^{-\lambda} - \frac{(1 - e^{-\lambda})}{r^2} = \kappa p_r ,\label{e1}
\end{equation}

\begin{equation}
-\kappa {T^2}_2 = -\kappa {T^3}_3 =
\left[\frac{{\nu}^{\prime\prime}}{2} -
\frac{{\lambda}^{\prime}{\nu}^{\prime}}{4} +
\frac{{{\nu}^{\prime}}^2}{4} + \frac{{\nu}^{\prime} -
{\lambda}^{\prime}}{2r}\right]e^{-\lambda} = \kappa p_t,\label{e2}
\end{equation}

\begin{equation}
\kappa {T^4}_4 = \frac{{\lambda}^{\prime}}{r} e^{-\lambda} +
\frac{(1 - e^{-\lambda})}{r^2} = \kappa \rho,\label{e3}
\end{equation}
where the prime denotes differential with respect to radial
coordinate $r$.

The pressure anisotropy condition for the system can be provided
as
\begin{equation}
\Delta=\kappa \left( p_{t}-p_{r}
\right)=\left[\frac{{\nu}^{\prime\prime}}{2} -
\frac{{\lambda}^{\prime}{\nu}^{\prime}}{4} +
\frac{{{\nu}^{\prime}}^2}{4} + \frac{{\nu}^{\prime} -
{\lambda}^{\prime}}{2r}\right]e^{-\lambda}-\frac{{\nu}^{\prime}}{r}
e^{-\lambda} + \frac{(1 - e^{-\lambda})}{r^2} .\label{e4}
\end{equation}

Now let us consider the metric potentials \cite{Durgapal1985} in
the following forms:
\begin{equation}
e^{-\lambda}=\frac{7-10Cr^2-C^2r^4}{7+14Cr^2+7C^2r^4},
\end{equation}

\begin{equation}
\nu=2\ln\psi,
\end{equation}
where $C$ is a positive constant and $\psi$ is a function which
depends on radial coordinate $r$. The nature of plots for these
quantities are shown in Fig. 1.

\begin{figure}[h]
\centering
\includegraphics[width=5cm]{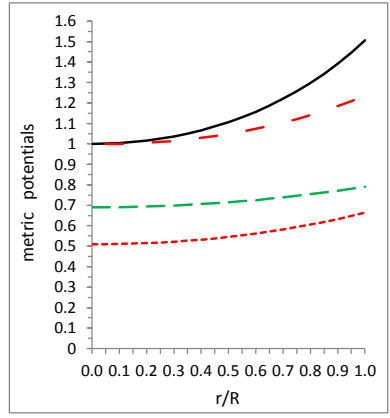}
\caption{Variation of the metric potentials with radial coordinate
$r/R$ are shown in the above figure. Here the legends are as
follows: (i) $e^{\nu}$ is plotted with dotted line for Her~X-1 and
short-dashed line for white dwarf, (ii) $e^{\lambda}$ is plotted
with continuous line for Her X-1 and long-dashed line for white
dwarf}
\end{figure}

The above Eq. (7) together with Eqs. (8) and (9) becomes
\begin{equation}
\Delta=\left[\frac{7-10Cr^2-C^2r^4}{7(1+Cr^2)^2}\right]
\frac{{\psi}^{\prime\prime}}{\psi}+\left[\frac{C^3r^6+19C^2r^4-21Cr^2-7}{7r(1+Cr^2)^3}\right]
\frac{{\psi}^{\prime}}{\psi}
+\left[\frac{8C^2r^2(Cr^2+5)}{7(1+C{r}^2)^3}\right].
\end{equation}

\begin{figure}[h]
\centering
\includegraphics[width=5cm]{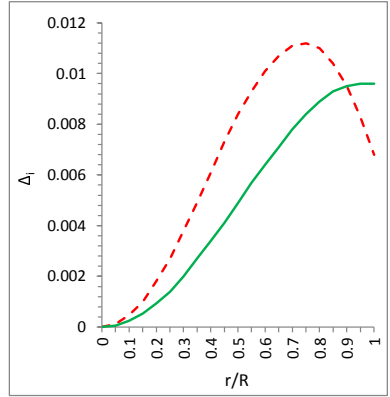}
\caption{Variation of the anisotropy factor with radial coordinate
$r/R$ are shown in figure. Here the legends are as follows: (i)
$\Delta$ is plotted with short-dashed line for Her X-1, (ii)
$\Delta$ is plotted with continuous line for white dwarf}
\end{figure}

\section{The solutions for the model}

Here our initial aim is to find out the pressure anisotropic
function $\Delta$, which is zero at the centre and monotonic
increasing for suitable choices of $\psi$. However, Lake
\cite{Lake2003} imposes condition that $\psi$ should be regular
and monotonic increasing function of radial coordinate $r$.

Let us therefore take the form of $\psi$ as follows:
\begin{equation}
\psi=(1-\alpha+Cr^2)^2,
\end{equation}
where $\alpha > 0$.

Substituting the value of $\psi$ from Eq. (11) in Eq. (10), we get
\begin{equation}
\Delta=-{\frac {8}{7}}\,{\frac {\alpha\,{C}^{2}{r}^{2}[2\,{C}^{2}{r}^{
4}+ \left( 16-\alpha \right) C{r}^{2}-5\,\alpha-2]}{ \left( 1+C{r}^{2}
 \right) ^{3} \left( 1-\alpha+C{r}^{2} \right) ^{2}}}.
\end{equation}

For $\alpha > 0$ and $0 < Cr^2 <
\frac{\sqrt{{\alpha}^{2}+8\alpha+272}-(16-K)}{4}$, the pressure
anisotropy is finite as well as positive everywhere as can be seen
in Fig. 2.

By inserting the above value of $\Delta$ in the Eq. (12), we get
\begin{eqnarray}
\psi^{\prime\prime}+\left[\frac{C^3r^6+19C^2r^4-21Cr^2-7}{r(1+Cr^2)(7-10Cr^2-C^2r^4)}\right]\psi^{\prime}
\nonumber \\ +
\frac{1}{7-10Cr^2-C^2r^4}\left[\frac{8C^2r^2(Cr^2+5)}{(1+Cr^2)}+
\frac{8\alpha
C^2r^2[2C^2r^4+(16-\alpha)Cr^2-(5\alpha+2)]}{(1+Cr^2)(1-\alpha+Cr^2)^2}\right]\psi=0.
\end{eqnarray}

Now our next task is to obtain the most general solution of the
differential Eq. (13). Here we shall use the change of dependent
variable method. We consider the differential equation of the form
\begin{equation}
y^{\prime\prime}+p(r)y^{\prime}+q(r)y=0.
\end{equation}

Let $y=y_{1}$ be the particular solution of the differential Eq.
(14). Then $y=y_{1}U$ will be complete solution of the
differential Eq. (14), where
$$U=a_{1}+b_{1}\int\exp\left[-\int(p(r)+\frac{2y^{\prime}}{y_{1}})dr\right]dr,$$
where $a_{1}$ and $b_{1}$ are arbitrary constants.

Again let us consider here that
$\psi=(1-\alpha+Cr^2)^2=\psi_{\alpha r}$ is a particular solution
of Eq. (13). So, the most general solution of the differential Eq.
(13) can be given by
\begin{equation}
\psi=(1-\alpha+Cr^2)^2\left[\tilde{B}+\tilde{A}\int\exp\left\{-\int\left(\frac{C^{3}r^{6}
+19C^{2}r^{4}-21Cr^{2}-7}{r(1+Cr^{2})(7-10Cr^{2}-C^{2}r^{4})}
+\frac{8Cr^{2}}{r(1-\alpha+Cr^2)}\right)dr\right\}dr\right],
\end{equation}
where $\tilde{A}$ and $\tilde{B}$ are arbitrary constants.

After integrating it, we get
\begin{equation}
\psi=\psi_{\alpha r}\left[B-A\left\{\frac{\{\psi_{\alpha
1}+\psi_{\alpha 2}(1-\alpha+Cr^2)+\psi_{\alpha 3}\psi_{\alpha
r}\}\sqrt{(\psi_{\alpha 5}-2(4+\alpha)(1-\alpha+Cr^2)
-\psi_{\alpha r})}}{(1-\alpha+Cr^2)^3} +W(r)\right\}\right],
\end{equation}
where
\begin{equation}
\psi_{\alpha r}=(1-\alpha+Cr^2)^2,
\end{equation}

\begin{equation}
W(r)=\frac{\psi_{\alpha 4}}{\sqrt{\psi_{\alpha 5}}}
\log\left[\frac{\psi_{\alpha 5}-(4+\alpha)(1-\alpha+Cr^{2})+
\sqrt{\psi_{\alpha 5}}\sqrt{(\psi_{\alpha
5}-2(4+\alpha)(1-\alpha+Cr^2)-\psi_{\alpha
r})}}{(1-\alpha+Cr^2)\psi_{\alpha 5}}\right],
\end{equation}
and $A$ and $B$ are arbitrary constants with

$\psi_{\alpha 1}=\frac{\alpha}{3(16-8\alpha-\alpha^{2})}$,\\

$\psi_{\alpha
2}=\frac{24-2\alpha+\alpha^{2}}{3(16-8\alpha-\alpha^{2})^{2}}$,\\

$\psi_{\alpha
3}=\frac{288+80\alpha-10\alpha^{2}+\alpha^{3}}{3(16-8\alpha-\alpha^{2})^{3}}$,\\

$\psi_{\alpha
4}=\frac{1536-384\alpha+48\alpha^{2}-2\alpha^{3}}{3(16-8\alpha-\alpha^{2})^{3}}$,\\

$\psi_{\alpha 5}=(16-8\alpha-\alpha^{2})$.\\

Using Eqs. (8), (12) and (16) the expressions for energy-density
and pressure read as
\begin{equation}
\frac{\kappa\rho}{C}=\frac{8(9+2Cr^{2}+C^{2}r^{4})}{7(1+Cr^{2})^{3}},
\end{equation}

\begin{figure}[h]
\centering
\includegraphics[width=5cm]{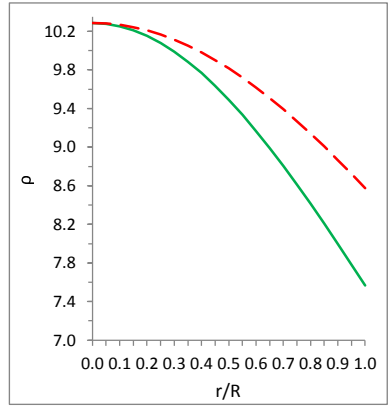}
\caption{Variation of the density with radial coordinate $r/R$ are
shown in figure. Here the legends are as follows: (i) $\rho$ is
plotted with continuous line for Her X-1 (ii) $\rho$ is plotted
with dashed line for white dwarf}
\end{figure}

\begin{figure}[h]
\centering
\includegraphics[width=5cm]{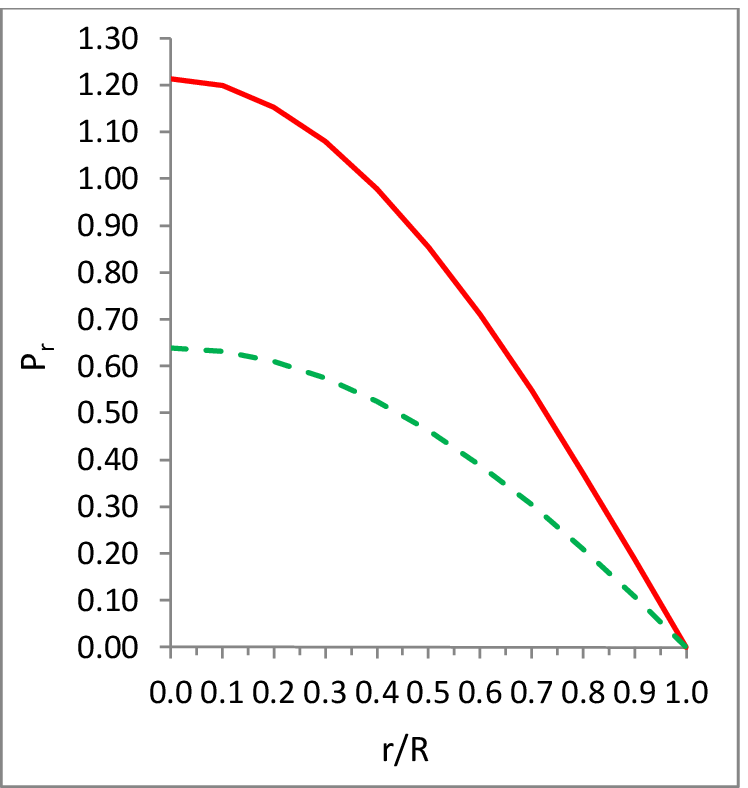}
\includegraphics[width=5cm]{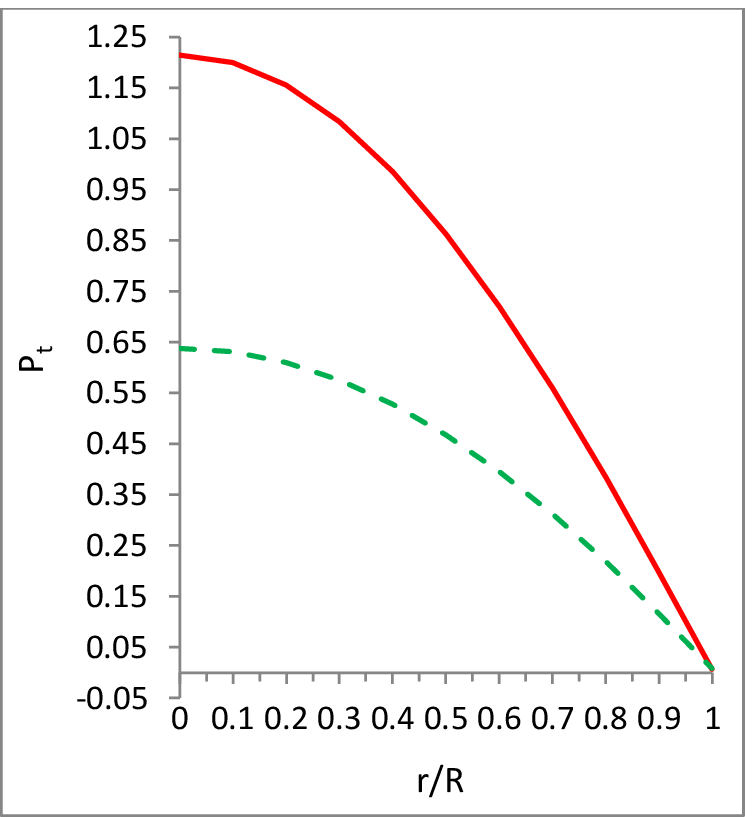}
\caption{Variation of the density with radial coordinate $r/R$ are
shown in figure. Here the legends are as follows: (i) $p_r$ is
plotted with continuous line for Her X-1 and short-dashed line for
white dwarf in left graph (left panel), (ii) $p_t$ is plotted with
continuous line for Her X-1 and short-dashed line for white dwarf
in right graph (right panel)}
\end{figure}

and

\begin{equation}
\frac{\kappa p_{r}}{C}=\frac{4(7-10Cr^2 -
C^{2}r^{4})}{7(1+Cr^{2})^2}\left[\frac{\psi_{pr}(1-\alpha+Cr^2)^{3}+2\psi}{\psi(1-\alpha+Cr^2)}\right]-
\frac{8(Cr^{2}+3)}{7(1+Cr^{2})^{2}},
\end{equation}
where
\begin{equation}
\psi_{pr}=\frac{A(1+Cr^{2})}{(1-\alpha+Cr^2)^{4}\sqrt{(7-10Cr^{2}-Cr^{2})}}.
\end{equation}

\begin{figure}[h]
\centering
\includegraphics[width=5cm]{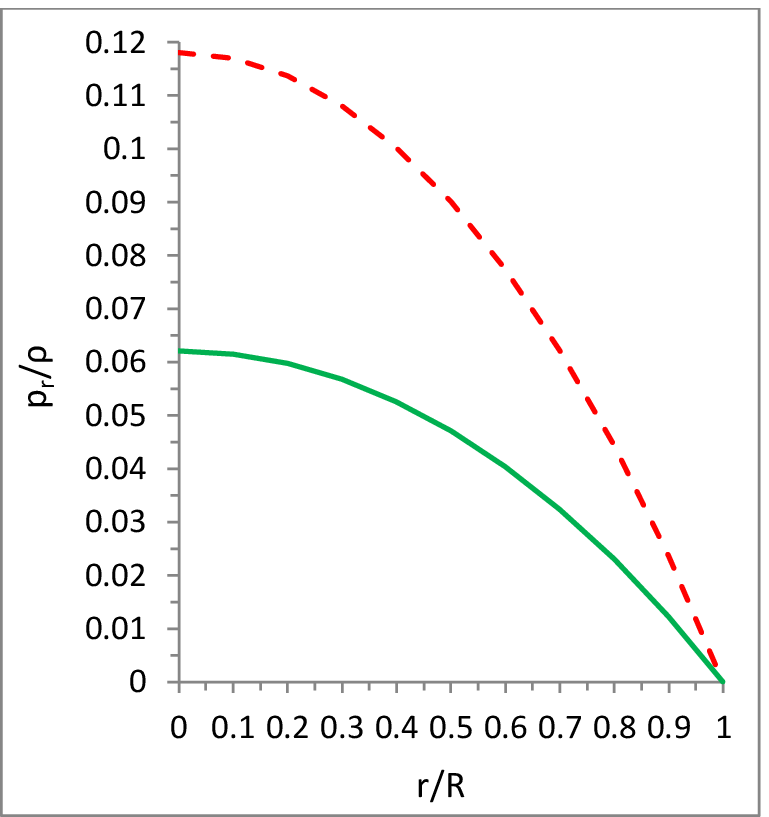}
\includegraphics[width=5cm]{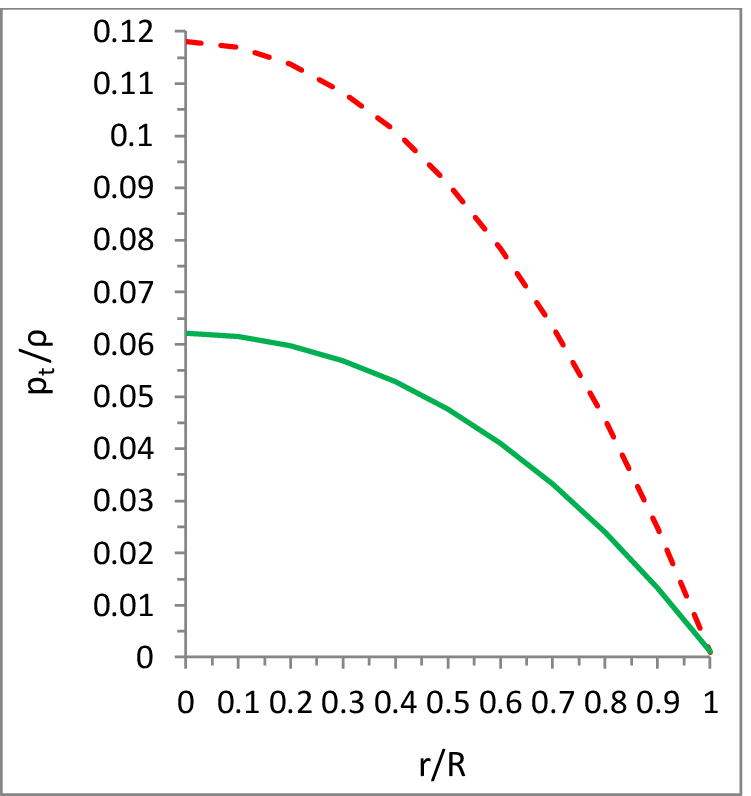}
\caption{Variation of the density with radial coordinate $r/R$ are
shown in figure. Here the legends are as follows: (i) $p_r/\rho$
is plotted with short-dashed line for Her X-1 and continuous line
for white dwarf (left panel), (ii) $p_t/\rho$ is plotted with
short-dashed line for Her X-1 and continuous line for white dwarf
(right panel) }
\end{figure}

In the Figs. 3-5 we have plotted the nature of the above physical
quantities which show viable features of the present model.

\subsection{Matching condition}
The above system of equations is to be solved subject to the
boundary condition that radial pressure $p_{r}=0$ at $r=R$ (where
$r=R$ is the outer boundary of the fluid sphere). It is clear that
$m (r=R)=M$ is a constant and, in fact, the interior metric (2.1)
can be joined smoothly at the surface of spheres ($r=R$), to an
exterior Schwarzschild metric whose mass is same as above i.e. $m
(r=R)=M$ \cite{Misner1964}.

The exterior spacetime of the star will be described by the
Schwarzschild metric given by
\begin{equation}
ds^2 = - \left(1-\frac{2M}{r}\right)^{-1}dr^2 - r^2(d\theta^2 +
sin^2\theta d\phi^2) + \left(1-\frac{2M}{r}\right) dt^2.
\end{equation}

Continuity of the metric coefficients $g_{tt}$, $g_{rr}$ across
the boundary surface $r=R$ between the interior and the exterior
regions of the star yields the following conditions:
\begin{equation}
\left(1-\frac{2M}{R}\right)^{-1} = e^{\lambda(R)},
\end{equation}

\begin{equation}
\left(1-\frac{2M}{r}\right) = \psi^{2}_R,
\end{equation}
where $\psi(r=R)=\psi_{R}$.

Equations (23) and (24) respectively give
\begin{equation}
M=\frac{R}{2}\left[\frac{8CR^{2}(3+CR^{2})}{7(1+CR^2)^2}\right],
\end{equation}

\begin{equation}
A=\frac{\sqrt{7-10CR^{2}-C^{2}R^{4}}}{\sqrt{7}(1+CR^{2})\psi_{\alpha
R}(\frac{B}{A}- \Omega(R))}.
\end{equation}

The radial pressure $p_{r}$ is zero at the boundary ($r=R$)
provides
\begin{equation}
\frac{B}{A}=\frac{(1+CR^{2})\sqrt{7-10CR^{2}-C^{2}R^{2}}}{2(1-\alpha+CR^{2})^{3}
[(1-\alpha+CR^{2})(3+CR^{2})-(7-10CR^{2}-C^{2}R^{4})]} \Omega(R),
\end{equation}
where
\begin{equation}
\psi_{\alpha R}=(1-\alpha+CR^{2})^{2},
\end{equation}

\begin{equation}
\Omega(R)=\frac{\{\psi_{\alpha 1}+\psi_{\alpha
2}(1-\alpha+CR^2)+\psi_{\alpha 3}\psi_{\alpha R}\}
\sqrt{\psi_{\alpha 5}-2(4+\alpha)(1-\alpha+CR^2)-\psi_{\alpha
R}}}{(1-\alpha+CR^{2})^3} + W(R),
\end{equation}

\begin{equation}
W(R)=\frac{\psi_{\alpha 4}}{\sqrt{\psi_{\alpha
5}}}\log\left[\frac{\psi_{\alpha 5}-(4+\alpha)(1-\alpha+CR^{2})+
\sqrt{\psi_{\alpha 5}}\sqrt{\psi_{\alpha
5}-2(4+\alpha)(1-\alpha+CR^{2})-\psi_{\alpha
R}}}{(1-\alpha+CR^{2})\psi_{\alpha 5}}\right].
\end{equation}

\section{Some physical features of the model}

\subsection{Regularity at centre}

The density $\rho$ and radial pressure $p_r$ and tangential
pressure $p_t$ should be positive inside the star. The central
density at centre for the present model is
\begin{equation}
\rho_{0}=\rho(r=0)=\frac{72C}{7}.
\end{equation}

The metric Eq. (22) implies that $C=\frac{7\rho_{0}}{72}$ is
positive finite.

Again, from Eq. (20), we obtain
\begin{equation}
\frac{p_{r}(r=0)}{C}=\frac{4A}{\sqrt7 (1-\alpha)^{2}\psi_{r=0}} -
\frac{24}{7},
\end{equation}
where $p_{r}(r=0)>0$.

This immediately implies that
\begin{eqnarray}
\frac{B}{A}<\frac{\sqrt{7}}{6(1-\alpha)^4} + \frac{\{\psi_{\alpha
1}+\psi_{\alpha 2}(1-\alpha)+\psi_{\alpha
3}(1-\alpha)^2\}\sqrt{\psi_{\alpha
5}-2(4+\alpha)(1-\alpha)-(1-\alpha)^2}}{(1-\alpha)^3} \nonumber
\\ + \frac{\psi_{\alpha 4}}{\sqrt{\psi_{\alpha 5}}}
\log\left[\frac{\psi_{\alpha
5}-(4+\alpha)(1-\alpha)+\sqrt{\psi_{\alpha 5}}\sqrt{\psi_{\alpha
5}-2(4+\alpha)(1-\alpha)-(1-\alpha)^2}}{(1-\alpha)\psi_{\alpha
5}}\right].
\end{eqnarray}

\subsection{Causality conditions}

Inside the fluid sphere the speed of sound should be less than the
speed of light i.e. $0 \leq V_{sr}=\sqrt{\frac{dp_{r}}{d\rho}}<1$
and $0 \leq V_{st}=\sqrt{\frac{dp_{t}}{d\rho}}<1$. Therefore
\begin{equation}
{V^{2}}_{sr}=(1+Cr^{2})\left[\frac{4(C^{2}r^{4}+10Cr^{2}-7)(1+Cr^{2})(\psi_{r1}
-\psi_{r2}-\psi_{r3})-8(Cr^{2}+5)}{8(C^{2}r^{4}+2Cr^{2}+25)}\right],
\end{equation}

\begin{figure}[h]
\centering
\includegraphics[width=5cm]{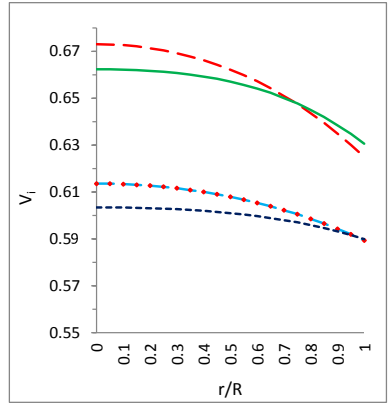}
\caption{Variation of the sound velocity with radial coordinate
$r/R$ are shown in figure. Here the legends are as follows: (i)
$V_r$ is plotted with dash line for Her X-1, (ii) $V_r$ is plotted
with marker continuous line for white dwarf (iii) $V_t$ is plotted
with continuous line for Her X-1 (iv) $V_t$ is plotted with dotted
line for white dwarf}
\end{figure}

\begin{figure}[h]
\centering
\includegraphics[width=5cm]{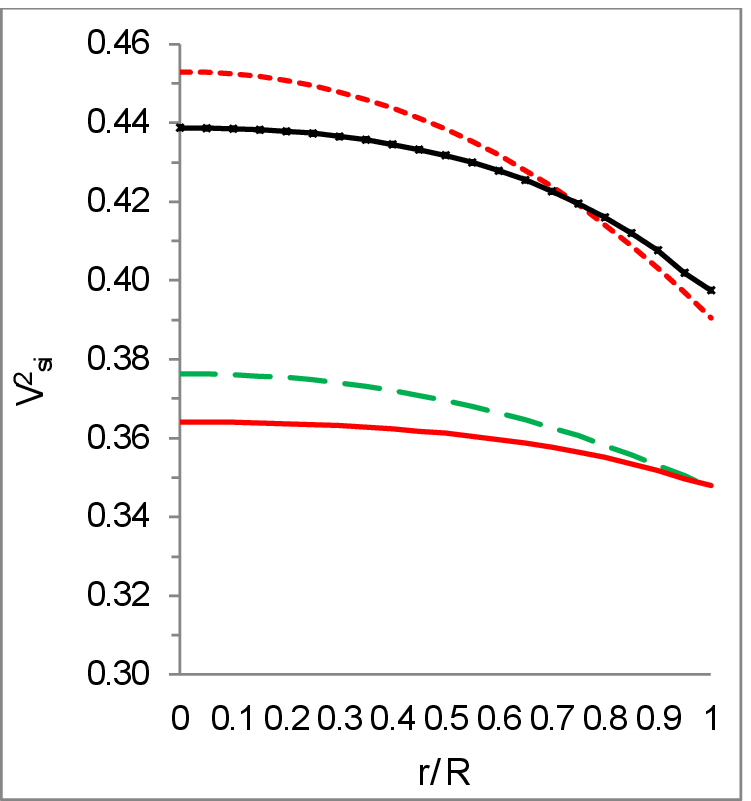}
\caption{Variation of the sound velocity with radial coordinate
r/R are shown in figure. Here the legends are as follows: (i)
$V^2_{sr}$ is plotted with dotted line for Her X-1, (ii)
$V^2_{sr}$ is plotted with dashed line for white dwarf (iii)
$V^2_{st}$ is plotted with marker continuous line for Her X-1 (iv)
$V^2_{st}$ is plotted with continuous line for white dwarf}
\end{figure}

\begin{equation}
{V^{2}}_{st}=(1+Cr^{2})\left[\frac{4(C^{2}r^{4}+10Cr^{2}-7)(1+Cr^{2})(\psi_{r1}
-\psi_{r2}-\psi_{r3})-8(Cr^{2}+5)-\psi_{r4}}{8(C^{2}r^{4}+2Cr^{2}+25)}\right],
\end{equation}
where
\begin{equation}
\psi_{pr}=\frac{A(1+Cr^{2})}{(1-\alpha+Cr^{2})^{4}\sqrt{(7-10Cr^{2}-Cr^{2})}},
\end{equation}

\begin{equation}
\psi_{r1}=[\frac{2}{(1-\alpha+Cr^{2})^{2}} +
\frac{4(3-Cr^{2})(1-\alpha+Cr^{2})^{2}\psi_{pr}}{\psi(1+Cr^{2})(7-10Cr^{2}-C^{2}r^{4})}],
\end{equation}

\begin{equation}
\psi_{r2}=[\frac{2K}{(1-\alpha+Cr^{2})} +
\frac{(1-\alpha+Cr^{2})^{2}\psi_{pr}}{\psi}]^{2},
\end{equation}

\begin{equation}
\psi_{r3}=\frac{8(3-Cr^{2})}{(7-10Cr^{2}-C^{2}r^{4})(1+Cr^{2})}[\frac{2}{(1-\alpha+Cr^{2})}
+ \frac{(1-\alpha+Cr^{2})^{2}\psi_{pr}}{\psi}],
\end{equation}

\begin{equation}
\psi_{r4}=[4\alpha\frac{(1+Cr^{2})(1-\alpha+Cr^{2})\psi_{r5} -
Cr^{2}\psi_{r6}(5-3\alpha+5Cr^{2})}{(1+Cr^{2})(1-\alpha+Cr^{2})^{3}}],
\end{equation}

\begin{equation}
\psi_{r5}=[12C^{2}r^{4}+4(16-\alpha)Cr^{2}-(10\alpha+4)],
\end{equation}

\begin{equation}
\psi_{r6}=[4C^{2}r^{4}+2(16-\alpha)Cr^{2}-(10\alpha+4)].
\end{equation}

The physical quantities related to the above equations are plotted
in Figs. 6 and 7.

\subsection{Well behaved condition}
The velocity of sound is monotonically decreasing away from the
centre and it is increasing with the increase of density i.e.
$\frac{d}{dr}(\frac{dp_{r}}{d\rho})<0$  or
$(\frac{d^{2}p_{r}}{d\rho^{2}})>0$  and
$\frac{d}{dr}(\frac{dp_{t}}{d\rho})<0$ or
$(\frac{d^{2}p_{t}}{d\rho^{2}})>0$ for $0 \leq r \leq R$. In this
context it is worth mentioning that the equation of state at
ultra-high distribution has the property that the sound speed is
decreasing outwards \cite{Canuto1973} as can be observed from Fig.
6.

\subsection{Energy conditions}
The anisotropic fluid sphere composed of strange matter will
satisfy the null energy condition (NEC), weak energy condition
(WEC) and strong energy condition (SEC), if the following
inequalities hold simultaneously at all points in the star:

NEC:  $\rho \geq 0$,

WEC:   $\rho + p_{r} \geq 0$,

WEC:  $\rho + p_{t} \geq 0$,

SEC:  $\rho + p_{r} +2p_{t} \geq 0$.

\begin{figure}[h]
\centering
\includegraphics[width=5cm]{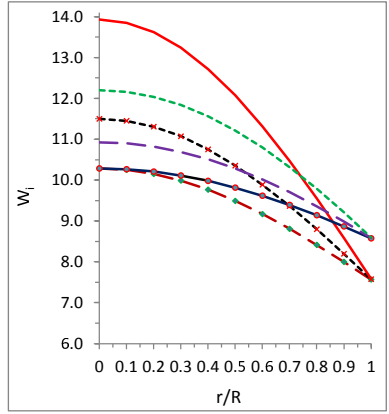}
\caption{Variation of the energy conditions with radial coordinate
$r/R$ are shown in figure. Here the legends are as follows: (i)
$NEC$ is plotted with marker long-dashed line, $WEC_r$ with marker
short-dashed line and $SEC$ with continuous line (for Her X-1),
(ii) $NEC$ is plotted with marker continuous line, $WEC_r$ with
long-dashed line and $SEC$ with short-dashed line (for white
dwarf)}
\end{figure}

We have shown the energy conditions in Fig. 8 for Her X-1 under
(i) and for white dwarf under (ii).

\subsection{Stability conditions}

\subsubsection{Case-1:} In order to have an equilibrium
configuration the matter must be stable against the collapse of
local regions. This requires Le Chatelier's principle, also known
as local or microscopic stability condition, that the radial
pressure $p_r$ must be a monotonically non-decreasing function of
$r$ such that $\frac{dp_{r}}{d\rho} \geq 0$ \cite{Bayin1982}.
Heintzmann and Hillebrandt \cite{Heintzmann1975} also proposed
that neutron star with anisotropic equation of state are stable
for $\gamma > 4/3$ as is observed from Fig. 9 and also shown in
Tables 1 and 2 of our model related to compact stars.

\begin{figure}[h]
\centering
\includegraphics[width=5cm]{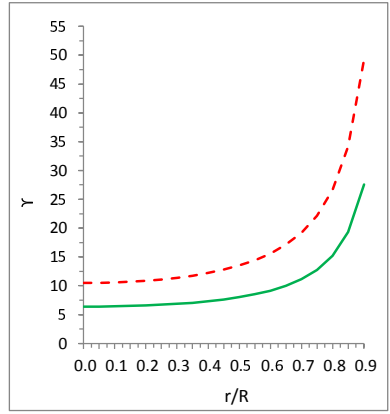}
\caption{Variation of the adiabatic index with radial coordinate
$r/R$ are shown in figure. Here the legends are as follows: (i)
$\gamma$ is plotted with continuous line for Her X-1 (ii) $\gamma$
is plotted with dash line for white dwarf}
\end{figure}

\subsubsection{Case-2:} For physically acceptable model, one expects that the
velocity of sound should be within the range
$0=V_{si}^2=(dp_{i}/d\rho) \leq 1 $ \cite{Herrera1992,Abreu2007}.
We plot the radial and transverse velocity of sound in Fig. 7 and
conclude that all parameters satisfy the inequalities $
0=V_{sr}^2=(dp_{i}/d\rho) \leq 1 $ and $0=V_{st}^2=(dp_{i}/d\rho)
\leq 1 $ everywhere inside the star models. Also $ 0=V_{st}^2 \leq
1 $ and $0=V_{sr}^2 \leq 1 $, therefore $|V_{st}^2-V_{sr}^2| \leq
1$. Now, to examine the stability of local anisotropic fluid
distribution, we follow the cracking (also known as overturning)
concept of Herrera \cite{Herrera1992} which states that the region
for which radial speed of sound is greater than the transverse
speed of sound is a potentially stable region.

For this we calculate the difference of velocities as follows:
\begin{equation}
V_{st}^2-V_{sr}^2=\alpha \left[
\frac{(1+Cr^2)(1-\alpha+Cr^2)\psi_{r5}-Cr^2\psi_{r6}(5-3\alpha+5Cr^2)}{2(1+Cr^2)^3(1-\alpha+Cr^2)^3(C^2r^4+2Cr^2+25)}\right],
\end{equation}
where
\begin{equation}
\psi_{r5}=[12C^{2}r^{4}+4(16-\alpha)Cr^{2}-(10\alpha+4)],
\end{equation}

\begin{equation}
\psi_{r6}=[4C^{2}r^{4}+2(16-\alpha)Cr^{2}-(10\alpha+4)].
\end{equation}

\begin{figure}[h]
\centering
\includegraphics[width=5cm]{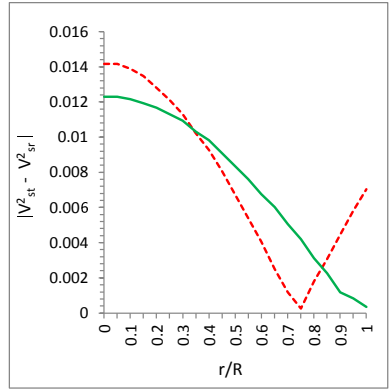}
\caption{Variation of the absolute value of square of sound
velocity with radial coordinate $r/R$ are shown in figure. Here
the legends are as follows: (i) $|V^2_{st} - V^2_{sr}|$ is plotted
with short-dashed line for Her X-1, (ii) $|V^2_{st} - V^2_{sr}|$
is plotted with continuous line for white dwarf}
\end{figure}

It can be seen that $|V_{st}^2-V_{sr}^2|$ at the centre lies
between 0 and 1 (see Fig. 10). This implies that we must have $0
\leq \frac{\alpha(10\alpha+4)}{50(1-\alpha)^2} \leq 1$. Then
$\alpha$ should satisfy the following condition: $0 \leq \alpha
\leq \frac{52-\sqrt{704}}{40}$.

\subsection{Generalized TOV equation}
The generalized Tolman-Oppenheimer-Volkoff (TOV) equation
\begin{equation}
-\frac{M_G\left(\rho+p_r\right)}{r^2}e^{\frac{\lambda-\nu}{2}}-\frac{dp_r}{dr}
+\frac{2}{r}\left(p_t-p_r\right)=0, \label{tov}
\end{equation}
where $M_G=M_G(r)$ is the effective gravitational mass which can
be given by
\begin{equation}
M_G(r)=\frac{1}{2}r^2e^{\frac{\nu-\lambda}{2}}\nu^{\prime}.
\end{equation}

Substituting the value of $M_G(r)$ in Eq. (46), we get
\begin{equation}
-\frac{1}{2}\nu^{\prime}(\rho+p_r)-\frac{dp_r}{dr}
+\frac{2}{r}\left(p_t-p_r\right)=0.
\end{equation}

Equation (48) basically describes the equilibrium condition for an
anisotropic fluid subject to gravitational ($F_g$), hydrostatic
($F_h$) and anisotropic stress ($F_a$) which can, in a compact
form, be expressed as
\begin{equation}
 F_g+ F_h+ F_a=0,
\end{equation}
where
\begin{equation}
F_{g}=-\frac{1}{2}\nu^{\prime}(\rho+p_r),
\end{equation}

\begin{equation}
F_{h}=-\frac{dp_r}{dr},
\end{equation}

\begin{equation}
F_{a}=\frac{2}{r}\left(p_t-p_r\right).
\end{equation}

The above forces can be expressed in the following explicit forms:
\begin{eqnarray}
F_{g}=-\frac{1}{2}\nu^{\prime}(\rho+p_r)=\frac{C^{2}r}{8\pi}\left[\frac{8(6-2Cr^2)}{7(1+Cr^2)^3}\frac{[\psi_{pr}(1-\alpha+Cr^2)^3+2\psi]}{\psi(1-\alpha+Cr^2)}
\right. \nonumber \\ + \left.
\frac{4(7-C^2r^4-10Cr^2)}{7(1+Cr^2)^2}\left(\frac{\psi_{pr}(1-\alpha+Cr^2)^3+2\psi}{\psi(1-\alpha+Cr^2)}\right)^2\right],
\end{eqnarray}

\begin{equation}
F_{h}=-\frac{dp_r}{dr}=\frac{C^{2}r}{4\pi}\left[\frac{4(C^2r^4+10Cr^2-7)}{7(1+Cr^2)^2}(\psi_{r1}-\psi_{r2}-\psi_{r3})-\frac{8(Cr^2+5)}{7(1+Cr^2)^3}\right],
\end{equation}

\begin{equation}
F_{a}=\frac{2}{r}\left(p_t-p_r\right)=\frac{C^2r}{\pi}\left[\frac{2\alpha[(5\alpha+2)-(16-\alpha)Cr^2-2C^2r^4]}{7(1+Cr^2)^3(1-\alpha+Cr^2)^2}\right].
\end{equation}

\begin{figure}[h]
\centering
\includegraphics[width=5cm]{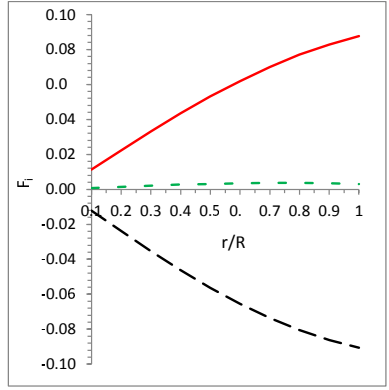}
\caption{Variation of the forces with radial coordinate $r/R$ are
shown in figure. Here the legends are as follows: $F_g$ is plotted
with long-dashed line, $F_h$ with continuous line and $F_a$ with
short-dashed line}
\end{figure}

Variation of different forces and attainment of equilibrium has
been drawn in Fig. 11.

\subsection{Effective mass-radius relation and surface redshift}

Let us now turn our attention towards the effective mass to radius
relationship. For static spherically symmetric perfect fluid star,
Buchdahl \cite{Buchdahl1959} has proposed an absolute constraint
on the maximally allowable mass-to-radius ratio ($M/R$) for
isotropic fluid spheres as $2M/R \leq 8/9$ (in the unit $c = G =
1$). This basically states that for a given radius a static
isotropic fluid sphere cannot be arbitrarily massive. However, for
more generalized expression for  mass-to-radius ratio one may look
at the paper by Mak and Harko \cite{Mak2003a}.

For the present compact star model, the effective mass is written
as
\begin{equation}
M_{eff}= 4\pi \int_0^R \rho
r^2dr=\frac{1}{2}R[1-e^{-\lambda(R)}]=\frac{1}{2}R\left[\frac{8CR^2(3+CR^2)}{7(1+2CR^2+C^2R^4)}\right].
\end{equation}

The compactness of the star is therefore can be given by
\begin{equation}
u=\frac{M_{eff}}{R}=\frac{1}{2}\left[\frac{8CR^2(3+CR^2)}{7(1+2CR^2+C^2R^4)}\right].
\end{equation}

\begin{figure}[h]
\centering
\includegraphics[width=5cm]{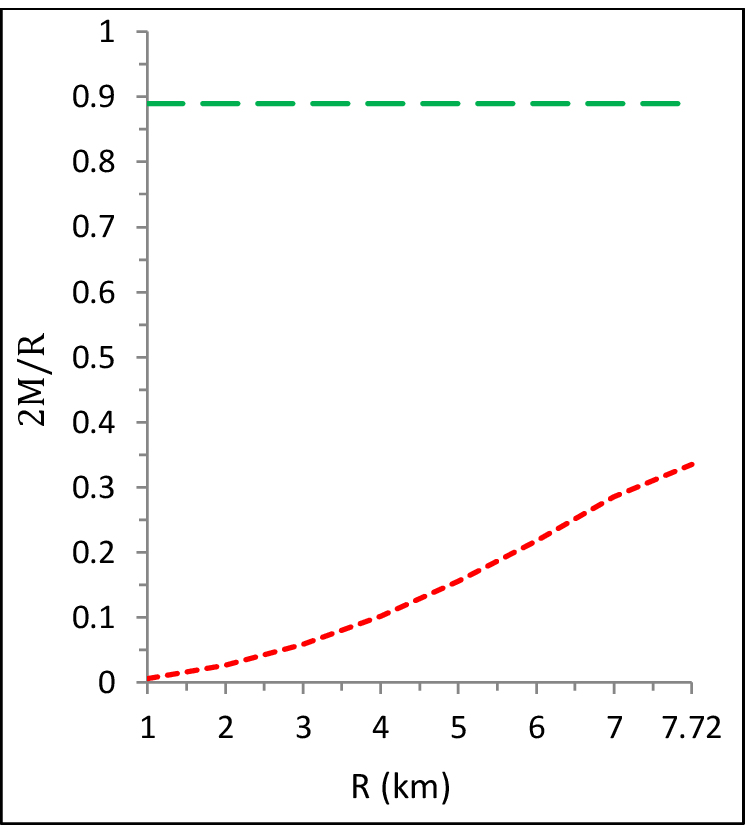}
\includegraphics[width=5cm]{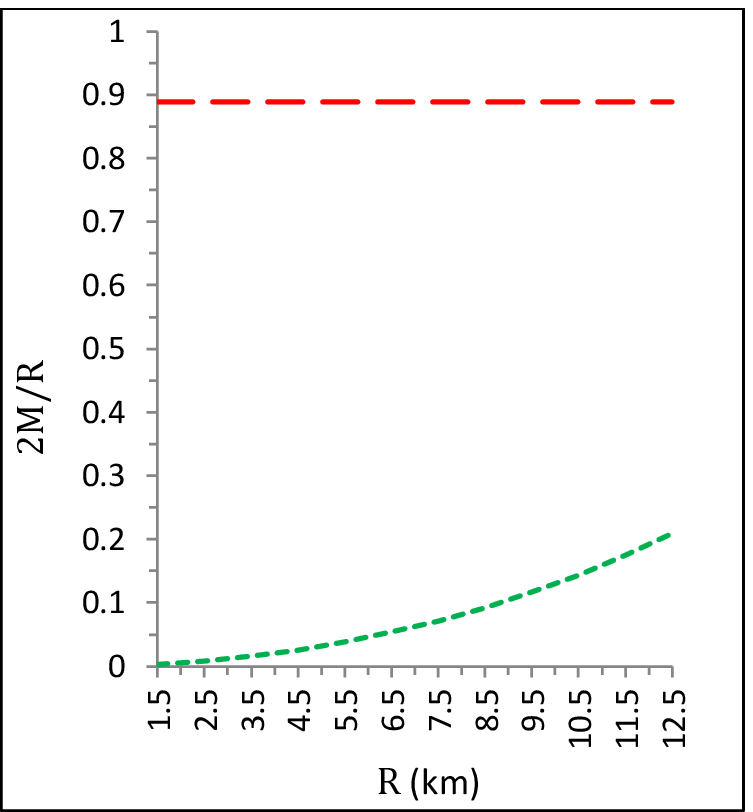}
\caption{Variation of the mass with radial coordinate $r/R$ are
shown in figure. Here the legends are as follows: (i) $2M/R$ is
plotted with short-dashed line for Her X-1 and long-dashed line
for its upper bound (left panel), (ii) $2M/R$ is plotted with
short-dashed line for white dwarf and long-dashed line for its
upper bound (right panel)}
\end{figure}

Therefore, the surface redshift ($Z$) corresponding to the above
compactness factor ($u$) is obtained as
\begin{equation}
Z=[1-2u]^{-1/2} - 1
=[1-\frac{8CR^2(3+CR^2)}{7(1+2CR^2+C^2R^4)}]^{-1/2} - 1.
\end{equation}

\begin{figure}[h]
\centering
\includegraphics[width=5cm]{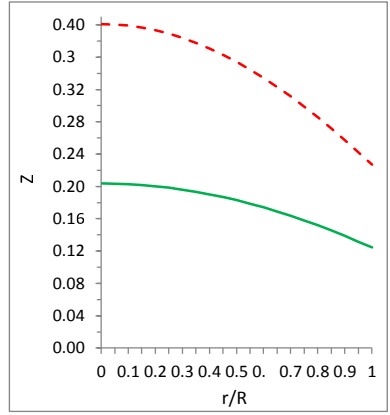}
\caption{Variation of the redshift index radial coordinate $r/R$
are shown in figure. Here the legends are as follows: $Z$ is
plotted with continuous line for Her X-1 and long-dashed line for
white dwarf}
\end{figure}

We have shown the variation of  physical quantities related to
Buchdahl's mass-to-radius ratio ($2M/R$) for isotropic fluid
spheres and also surface redshift are plotted in Figs. 12 and 13.

\section{Model parameters and comparison with some of the compact stars}

In this Section we prepare several data sheets for the model
parameters in the following Tables 1-3 and compare those with some
of the compact stars, e.g. Strange star Her X-1 and White dwarf in
Table 4. In our present investigation we propose a stable model
with the parameters $R=12.5202$~Km and $M=0.8882~M_{\odot}$ (for
white dwarf) whereas $R=7.7214$~Km and $M=0.8804$~$M_{\odot}$ (Her
X-1) type compact star. The values of these data points have
already been used for plotting graphs in the previous Sections 3
and 4 (See Figs. 1-13) in some way or others.

\begin{table}[h]
\centering \caption{Values of different physical parameters of
White dwarf star for
\newline ~$\alpha=0.10$,~$CR^2=0.068$,~$M=0.8882$~$M_{\odot}$,~$R=12.5202$~Km}\label{tbl-1}
\begin{tabular}{@{}lrrrrrrrrr@{}}
\hline

$r$     & $p_r$     & $p_t$  & $\rho$  & $V_r$  & $V_t$ & $\Delta$
& Z & $\gamma$\\ \hline

0.0 &0.6386  &0.6386  &10.2857 &0.6135  &0.6034  &0.00000 &0.2036
& 10.4949\\

0.1 &0.6314  &0.6316  &10.2663 &0.6133  &0.6033 &0.00024 &0.2028
&10.5863\\

0.2 &0.6096  &0.6106  &10.2084 &0.6127  &0.6031 &0.00093 &0.2003
&10.8718\\

0.3 &0.5738  &0.5758  &10.1129 &0.6116  &0.6026 &0.0020  &0.1961
&11.3897\\

0.4 &0.5245  &0.5278  &9.9814  &0.6100  &0.6019 &0.0034  &0.1903
&12.2191\\

0.5 &0.4624  &0.4673  &9.8157  &0.6079  &0.6010 &0.0049  &0.1830
&13.5139\\

0.6 &0.3885  &0.3949  &9.6185  &0.6053  &0.5997 &0.0064  &0.1741
&15.5930\\

0.7 &0.3040  &0.3118  &9.3926  &0.6022  &0.5980 &0.0078  &0.1637
&19.2098\\

0.8 &0.2101  &0.2190  &9.1412  &0.5985  &0.5959 &0.0089  &0.1519
&26.6355\\

0.9  &0.1083 &0.1178  &8.8676  &0.5942  &0.5932 &0.0095  &0.1388
&49.2412\\

1.0 &0.0000  &0.0096  &8.5754  &0.5893  &0.5899 &0.0096  &0.1244 &
$\infty$\\ \hline
\end{tabular}
\end{table}

\begin{table}[h]
\centering \caption{Values of different physical parameters of
Strange star Her X-1 for
\newline ~$\alpha=0.11$,~$CR^2=0.1178$,~$M=0.8804$~$M_{\odot}$,~$R=7.7214$~Km}\label{tbl-2}
\begin{tabular}{@{}lrrrrrrrrr@{}}
\hline

$r$     & $p_r$     & $p_t$  & $\rho$  & $V_r$  & $V_t$ & $\Delta$
& Z & $\gamma$\\ \hline

0.0 & 1.2135 & 1.2135 &10.2857 &0.6730  &0.6624 &0.0000  &0.4010
&4.2917\\

0.1 &1.1984  &1.1988 &10.2521 &0.6726  &0.6622 &0.0004 &0.3991
&4.3223\\

0.2 &1.1533 &1.1551 &10.1523 &0.6713 & 0.6617 &0.0018 &0.3933
&4.4180\\

0.3 &1.0795 &1.0833 &9.9890 &0.6692 &0.6607 &0.0038 &0.3838
&4.5919\\

0.4  &0.9790 &0.9851 &9.7665 &0.6662 &0.6592 &0.0061 &0.3707
&4.8710\\

0.5 &0.8546 &0.8630 &9.4906 &0.6622 &0.6571 &0.0084 &0.3541
&5.3078\\

0.6 &0.7096 &0.7197 &9.1681 &0.6572 &0.6541 &0.0101 &0.3342
&6.0116\\

0.7 &0.5475 &0.5586 &8.8067 &0.6510 &0.6501 &0.0111 &0.3113
&7.2403\\

0.8 &0.3725 &0.3835 &8.4143 &0.6436 &0.6450 &0.0110 &0.2856
&9.7716\\

0.9 &0.1886 &0.1982 &7.9990 &0.6350 &0.6385 &0.0095 &0.2574
&17.4995\\

1.0 &0.0000 &0.0068 &7.5686 &0.6249 &0.6305 &0.0068 &0.2271& -\\
\hline
\end{tabular}
\end{table}

\begin{table}[h]
\centering \caption{Values of the model parameters $A$, $B$, $C$
and $\alpha$ for different compact stars}\label{tbl-3}
\begin{tabular}{@{}lrrrrrrrrrrrrr@{}} \hline
Compact star    &$M$    & $R$   & $A$  & $B$  & $C$ & $\alpha$ \\
candidates & ($M_{\odot}$) & (Km)\\ \hline

White dwarf     & 0.8882    & 12.5202   & -2.1463    & 0.5533 &
$4.3380 \times 10^{-13}$ & 0.10\\

$Her~X-1$       & 0.8804     & 7.7214    & -1.6255 & 0.5301 &
$1.9758 \times 10^{-13}$ & 0.11\\ \hline
\end{tabular}
\end{table}

\begin{table}[h]
\centering \caption{Energy densities, central pressure and
Buchdahl condition  \newline for different compact star candidates
for the above parameter values of Tables 1 - 3}\label{tbl-4}
\begin{tabular}{@{}lrrrrrrrrr@{}}
\hline

Compact star        & Central Density           & Surface density
& Central pressure & Buchdahl condition\\

candidates          & ($gm/cm^{-3}$)            & ($gm/cm^{-3}$) &
($dyne/cm^{-2}$) & ($2M/R \leq 8/9$)\\ \hline

White dwarf       & $2.3961 \times 10^{14}$   & $2.0  \times
10^{14}$ & $1.3392  \times 10^{34}$ & 0.1418\\

$Her~X-1$           & $1.0913  \times 10^{15}$  & $0.8031 \times
10^{15}$ & $1.1591 \times 10^{35}$ & 0.2280\\ \hline
\end{tabular}
\end{table}

Practically what we have done in the tables are as follows: In
Tables 1-3 values of different physical parameters of Strange star
Her X-1 and White dwarf have been provided. Under this data set
then we calculate some physical parameters of compact star, say
central density, surface density, central pressure etc in Table 4.
It can be observed that these data are quite satisfactory for the
compact stars whether it is strange star with central density
$1.0913 \times 10^{15}$ $gm/cm^{-3}$ or white dwarf with central
density $2.3961 \times 10^{14}$ $gm/cm^{-3}$. Likewise this
feature of compact stars can be explored for some other physical
parameters also.

\section{Discussion and Conclusion}
In the present work we have investigated about an anisotropic
analogue of Durgapal-Fuloria \cite{Durgapal1985} and possibilities
of interesting physical properties of the proposed model. As a
necessary step we have contracted the anisotropic factor $\Delta$
by the help of both metric potentials $e^{\nu}$ and $e^{\lambda}$.
However, $e^{\lambda}$ is considered here same as
Durgapal-Fuloria~\cite{Durgapal1985} whereas $e^{\nu}$ is that
given by Lake~\cite{Lake2003}.

The field equations are solved by the change of dependent variable
method and under suitable boundary condition the interior metric
(2.1) has been joined smoothly at the surface of spheres ($r=R$),
to an exterior Schwarzschild metric whose mass is same as
m$(r=R)=M$ \cite{Misner1964}. The solutions set thus obtained are
correlated with the physical properties of some of the compact
stars which include strange star as well as white dwarf. It is
observed that the model is viable in connection to several
physical features which are quite interesting and acceptable as
proposed by other researchers within the framework of General
Theory of Relativity.

As a detailed discussion we would like to put forward here that
several verification scheme of the model have been performed and
extract expected results some of which are as follows:

(1) Regularity at centre: The density $\rho$ and radial pressure
$p_r$ and tangential pressure $p_t$ should be positive inside the
star. It is shown that the central density at centre is
$\rho_{0}=\rho(r=0)=\frac{72C}{7}$ and $p_{r}(r=0)>0$. This means
that the density $\rho$ as well as radial pressure $p_r$ and
tangential pressure $p_t$ all are positive inside the star.

(2) Causality conditions: It is shown that inside the fluid sphere
the speed of sound is less than the speed of light i.e. $0 \leq
V_{sr}=\sqrt{\frac{dp_{r}}{d\rho}}<1$, \quad $0 \leq
V_{st}=\sqrt{\frac{dp_{t}}{d\rho}}<1$.

(3) Well behaved condition: The velocity of sound is monotonically
decreasing away from the centre and it is increasing with the
increase of density as can be observed from Fig. 6.

(4)  Energy conditions: From Fig. 9 we observe that the
anisotropic fluid sphere composed of strange matter satisfy the
null energy condition (NEC), weak energy condition (WEC) and
strong energy condition (SEC) simultaneously at all points in the
star.

(5) Stability conditions: Following Heintzmann and Hillebrandt
\cite{Heintzmann1975} we note that neutron star with anisotropic
equation of state are stable for $\gamma > 4/3$ as is observed in
Tables 1 and 2 of our model. Also, it is expected that the
velocity of sound should be within the range
$0=V_{si}^2=(dp_{i}/d\rho) \leq 1 $ \cite{Herrera1992,Abreu2007}.
The plots for the radial and transverse velocity of sound in Fig.
7 everywhere inside the star models.

(6) Generalized TOV equation: The generalized
Tolman-Oppenheimer-Volkoff equation describes the equilibrium
condition for the anisotropic fluid subject to gravitational
($F_g$), hydrostatic ($F_h$) and anisotropic stress ($F_a$). Fig.
8 shows that the gravitational force is balanced by the joint
action of hydrostatic and anisotropic forces to attain the
required stability of the model. However, effect of anisotropic
force is very less than the hydrostatic force.

(7) Effective mass-radius relation and surface redshift: For
static spherically symmetric perfect fluid star, the Buchdahl
\cite{Buchdahl1959} absolute constraint on the maximally allowable
mass-to-radius ratio ($M/R$) for isotropic fluid spheres as $2M/R
\leq 8/9 = 0.8888$ is seen to be maintained in the present model
as can be observed from the Table 4.

In Sec. 5 we have made a comparative study by using model
parameters and data of two of the compact stars which are, in
general, very satisfactory as compared to the observational
results. However, at this point we would like to comment that the
sample data used for verifying the present model are to be
increased to obtain more satisfactory and exhaustive features in
the realm of physical reality.

\section*{Acknowledgement}
SKM acknowledges support from the Authority of University of
Nizwa, Nizwa, Sultanate of Oman. Also the author SR is thankful to
the authority of Inter-University Centre for Astronomy and
Astrophysics, Pune, India for providing him Associateship
programme under which a part of this work was carried out.

\end{document}